\documentclass{article}
\usepackage{ijcai24}

\usepackage{times}
\usepackage{soul}
\usepackage{url}
\usepackage[hidelinks]{hyperref}
\usepackage[utf8]{inputenc}
\usepackage[small]{caption}
\usepackage{graphicx}
\usepackage{amsmath}
\usepackage{amsthm}
\usepackage{booktabs}
\usepackage{algorithm}
\usepackage{algorithmic}
\usepackage[switch]{lineno}
\usepackage{enumitem}
\usepackage{amsfonts}
\usepackage{multirow}

\title{Incremental Learning of Stock Trends via Meta-Learning \\with Dynamic Adaptation}

\author{
	Shiluo Huang$^1$ \and
	Zheng Liu$^2$ \and
	Ye Deng$^1$\And
	Qing Li\footnote{Corresponding Author}$^{1}$
	\affiliations
	$^1$Southwestern University of Finance and Economics, Chengdu, China\\
	$^2$Tsinghua University, Beijing, China\\
}

\begin{document}
	
	\maketitle
	
	\begin{abstract}
		Forecasting the trend of stock prices is an enduring topic at the intersection of finance and computer science. Periodical updates to forecasters have proven effective in handling concept drifts arising from non-stationary markets. However, the existing methods neglect either emerging patterns in recent data or recurring patterns in historical data, both of which are empirically advantageous for future forecasting. To address this issue, we propose meta-learning with dynamic adaptation (MetaDA) for the incremental learning of stock trends, which periodically performs dynamic model adaptation utilizing the emerging and recurring patterns simultaneously. We initially organize the stock trend forecasting into meta-learning tasks and train a forecasting model following meta-learning protocols. During model adaptation, MetaDA efficiently adapts the forecasting model with the latest data and a selected portion of historical data, which is dynamically identified by a task inference module. The task inference module first extracts task-level embeddings from the historical tasks, and then identifies the informative data with a task inference network. MetaDA has been evaluated on real-world stock datasets, achieving state-of-the-art performance with satisfactory efficiency.
	\end{abstract}
	
	\section{Introduction}
	Stock trend forecasting is a classical but challenging problem, drawing the attention of both economists and computer scientists \cite{stockpredFin,multipattern}. However, the stock market is a recognized non-stationary environment where data distribution varies over time \cite{nonstatstock,stockus}. Such a phenomenon of distribution shift or concept drift \cite{stockdis,coneptdrift} presents a great challenge to the stock forecasting models. Meanwhile, stock data are typical streaming data: new stock data are generated as stock trading is conducted, resulting in an ever-growing dataset. Periodical model adaptation based on the latest datasets could prepare the forecaster for future tasks and alleviate the model aging issue \cite{DDGDA,Dadpt}.
	Up to now, there are two commonly used paradigms of model adaptation: Rolling Retraining (RR) \cite{DDGDA} and Incremental Learning (IL) \cite{Dadpt}. 
	
	RR retrains the model periodically, where a model is retrained with all the data in hand. In this way, RR could retain recurring patterns within the historical data \cite{recurrConcept}. The RR method focusing on predictable concept drift has achieved competitive performance in forecasting stock price trends \cite{DDGDA}, which empirically indicates the existence of recurring patterns in the stock data. However, the latest samples are reserved for the validation set in RR, which limits the model’s access to emerging patterns within the latest data. Meanwhile, it is straightforward to find that RR will become increasingly computation-consuming as the size of the dataset grows. 
	
	In contrast with RR, IL fine-tunes (adapts) the forecasting model with the latest data. The existing IL method \cite{Dadpt} can be viewed as a combination of meta-learning \cite{maml} and online learning \cite{onlineml,onlineGD}, where data is organized into meta-learning tasks. Compared with RR, IL efficiently adapts the model with a small number of gradient steps using incremental data. The performance of IL methods is competitive, which indicates the significance of emerging patterns within the latest data \cite{cdunpred,cdlatest}. However, the above model adaptation procedure disregards the historical data. As a domain-incremental problem \cite{threeincre}, stock trend forecasting still suffers from catastrophic forgetting \cite{cf} due to the existence of recurring patterns. The exclusion of historical data could therefore limit the performance of existing IL methods. 
	
	In short, the existing methods either fail to explicitly retrain the forecasting model with emerging patterns in the latest data or disregard the recurring patterns within historical data.
	To this end, we propose meta-learning with dynamic adaptation (MetaDA) for incremental learning of stock trends, which performs dynamic model adaptation considering both the recurring and emerging patterns. The adaptation efficiency of the proposed method is comparable to that of IL methods, as only the informative historical data is included in model adaptation. MetaDA organizes the data generated between two model adaptations into a set, and the sets are further transformed into meta-learning tasks. The target of model adaptation is to boost the performance on the upcoming data. We introduce a task inference network to estimate the patterns of upcoming stock data, and identify the historical data containing recurring patterns to complement the latest training data. To transform the sets of data into learnable pattern features, we also propose a task embedding method that extracts task-level representations from individual stock data. During model adaptation, a task-level feature is derived from a lookback window and then processed by the task inference network for the estimation of future patterns. A historical task that has the most similar pattern to the estimated pattern will be selected. The selected historical data will be used for model adaptation, along with the latest arriving training set. The proposed method could therefore exploit the beneficial patterns within both the latest data and the historical data. Experiments on the real-world stock datasets indicate that the proposed method has achieved state-of-the-art performance compared with existing IL and RR methods.
	The highlights of this paper are listed as follows.
	\begin{itemize}[leftmargin=8pt, labelsep = 3pt, topsep=0pt, parsep=0pt]
		\item As far as we know, MetaDA is the first method to incorporate both emerging and historical data into the incremental learning of stock trends.
		\item We propose a task inference module for the selection of informative historical data, which predicts the overall pattern of the subsequent data.
		\item The proposed method is evaluated on real-world datasets, and achieves competitive performance compared with the state-of-the-art methods.
	\end{itemize}
	
	\section{Related Work}
	\textbf{Concept Drifts in Stock Data.} Financial time series including stock data are inherently un-stationary, as the distribution of financial data changes over time \cite{conceptdriftoverview,finDS}. Such changes in the relation between the input features and target variable are usually known as concept drift \cite{csacm} or distribution shift \cite{dsnips}, which can be formally defined as follows.
	\begin{equation}
		\exists \mathbf{X}:\quad P_{t_1}(\mathbf{X},y) \neq P_{t_2}(\mathbf{X},y),
	\end{equation}
	where $ P_{t_1}(\mathbf{X},y)$ stands for the joint distribution of input $\mathbf{X}$ and target $y$, at time $t_1$. The concept drifts of stock data have been visualized via t-SNE \cite{tsne}, and indicated in Figure \ref{taskseg}. One of the commonly used assumptions in the concept drift community is that the concept drift is unpredictable (random). Studies following the above unpredictable assumption rely on the latest data for model adaptation \cite{conceptdriftlatest}. In stock trend forecasting, an incremental learning method that fine-tunes the model using the latest data exclusively has achieved state-of-the-art results \cite{Dadpt}. However, the empirical results of recent studies \cite{DDGDA} indicate that predictable (or recurring) concept drifts exist in stock data. It seems that the predictable and unpredictable concept drifts exist in the stock time series simultaneously. 
	
	\vspace{3.5pt}
	\noindent\textbf{Stock Trend Forecasting Against Concept Drift.} Up to now, there are only limited studies focused on handling the concept drifts of stock data. DoubleAdapt \cite{Dadpt} explores both the data adaptation and model adaptation for the incremental learning of stock data, which transforms incremental tasks into a bi-level optimization problem in the paradigm of meta-learning. But DoubleAdapt fails to handle the predictable concept drifts, as it adapts the model only with the latest data.
	DDG-DA \cite{DDGDA} focuses on predictable concept drifts. DDG-DA estimates the data distribution of the next time step and then generates training samples based on historical data. As an RR method, DDG-DA utilizes the recent data as the validation set and fails to update the forecasting model with the latest data. To the best of our knowledge, the previous studies have seldom incorporated both the latest and historical data to handle predictable and unpredictable concept drifts simultaneously.
	
	\section{Preliminaries}
	\textbf{Stock Price Trend Forecasting.} The price trend of stock $i$ at time $t$ is defined as the ratio of price change to the stock price at time $t$ \cite{REST,Dadpt}.
	\begin{equation}
		r^t_i = \frac{price^{t+1}_i - price^t_i}{price^t_i},
	\end{equation}
	where $price^t_i$ is the stock price at time $t$ and $r^t_i$ is the price trend. It should be noted that $price^t_i$ could be the closing price, opening price, and volume-weighted price.
	
	Denoting the feature (maybe factors) of stock $i$ at time $t$ as $\mathbf{x}^t_i \in \mathbb{R}^{1\times d}$, stock trend forecasters aim at predicting the price trend $r^t_i$ based on $\mathbf{x}^t_i$. In practice, $\mathbf{x}^t_i$ could involve a collection of indicators in the past few days, like opening price, closing price, and trading volume. Supposing there are $n$ observed stocks, the features of stocks form a data matrix $\mathbf{X}^t \in \mathbb{R}^{n\times d}$ at time $t$, while the stock trends at time $t$ can be summarized in a target vector $\mathbf{\Gamma}^t \in \mathbb{R}^{n\times 1}$. Stock forecasters learn the historical data $\{(\mathbf{X}^i,\mathbf{\Gamma}^i)\}_{i=1}^{t}$, and predict future stock trends $\{\mathbf{\Gamma}^i\}_{i=t+1}^{t+\tau}$ with upcoming stock features $\{\mathbf{X}^i\}_{i=t+1}^{t+\tau}$.
	
	\begin{figure}
		\centering
		\includegraphics[width=0.43\textwidth]{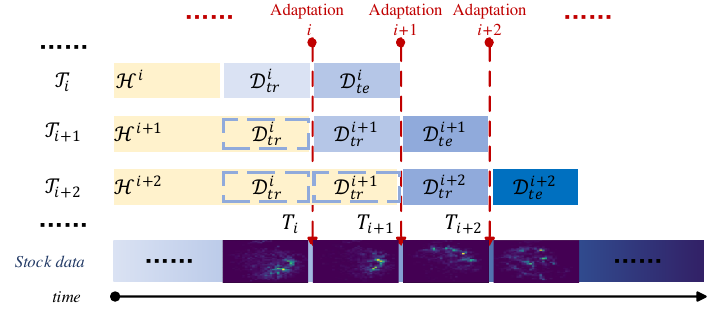}
		\caption{The incremental learning of stock trends, where the forecasting model is periodically updated via model adaptation. In the $i$ th model adaptation, the latest labeled data (training set) $\mathcal{D}_{tr}^i$, the upcoming data (testing set) $\mathcal{D}_{te}^i$, and the historical data $\mathcal{H}^i$ form the $i$ th learning task $\mathcal{T}_i$. We aim to improve the overall performance on testing sets across future tasks. The stock data are also visualized via t-SNE and the results are presented at the bottom of the figure.}
		\label{taskseg}
		\vspace{-0.3em}
	\end{figure}
	
	\vspace{3.5pt}
	\noindent \textbf{Incremental Learning of Stock Trends.} As shown in Figure \ref{taskseg}, the stock data is divided into segments based on model adaptation intervals, forming a sequence of meta-learning tasks. In this paper, a meta-learning task consists of the training set, the testing set, and the historical data. Let $\mathcal{T}_i = \{\mathcal{D}^i_{tr},\mathcal{D}^i_{te}, \mathcal{H}^i\}$ denote the $i$ th learning task. $\mathcal{D}^i_{tr}$ and $\mathcal{D}^i_{te}$ are the training and testing sets respectively, while $\mathcal{H}^i$ stands for the historical data. The aim of $i$ th model adaptation is to boost the prediction performance on $\mathcal{D}^i_{te}$ based on $\mathcal{D}^i_{tr}$ and $\mathcal{H}^i$. For the $i$ th model adaptation, $\mathcal{D}_{tr}^{i}$ consists of the stock data generated between the $i-1$ th and the $i$ th model adaptation, while testing set $\mathcal{D}^i_{te}$ contains the stock features generated between the $i$ th and $i+1$ th model adaptation.  Assuming that the date of $i$ th model adaptation is $T_i$, the testing set can be denoted as $\mathcal{D}^i_{te} =\{\mathbf{X}^j\}_{j=T_i}^{T_{i+1}-1}$. After collecting the true labels (price trends) of $\mathcal{D}^i_{te}$, $\mathcal{D}^i_{te}$ and the corresponding labels serve as the training set of $i+1$ th task, namely $\mathcal{D}^{i+1}_{tr}=\{(\mathbf{X}^j,\mathbf{\Gamma}^j)\}_{j=T_i}^{T_{i+1}-1}$. The training set of the former task will be collected into historical data $\mathcal{H}^{i+1}={\{\mathcal{D}^{j}_{tr}}\}_{j=1}^{i}$.
	
	\vspace{3.5pt}
	\noindent \textbf{Problem Formulation.} Considering an adaptation interval of $T_{ada}=T_{i+1} – T_i$, the stock trend forecasting is divided into a sequence of model adaptation tasks $\mathbb{\mathcal{T}}=\{\mathcal{T}_1,\mathcal{T}_2,...,\mathcal{T}_i,...\}$ with $\mathcal{T}_i = \{\mathcal{D}^i_{tr},\mathcal{D}^i_{te}, \mathcal{H}^i\}$. Given a forecasting model $F(\cdot, \Theta)$ with parameter $\Theta$, $F(\cdot, \Theta)$ is periodically fine-tuned following the model adaptation process. In the $i$ th task, the forecasting model $F(\cdot, \Theta)$ is adjusted with $\{\mathcal{D}^i_{tr}, \mathcal{H}^i\}$, then inferences the price trends of upcoming data $\mathcal{D}^i_{te}$, and finally learns the latest data after the arrival of true labels. We aim to achieve the highest forecasting performance across all the model adaptation tasks. 
	
	As both the predictable and un-predictable concept drifts exist in stock trend forecasting, updating the model with the historical ($\mathcal{H}$) or latest ($\mathcal{D}_{tr}$) data exclusively might lead to domain gap between the adaptation data and the upcoming data. Accordingly, model adaptation considering the informative information within both $\mathcal{H}$ and $\mathcal{D}_{tr}$ is likely to boost the forecasting performance.
	
	\section{MetaDA}
	The proposed method follows a training-adapting procedure: MetaDA first learns the labeled data in hand and then dynamically adapts the forecasting model $F(\cdot, \Theta)$ for the upcoming concept drifts. Given the adaptation interval $T_{ada}$, both the training and adapting data are organized into a list of tasks $\mathcal{T}$. The forecasting model $F(\cdot, \Theta)$ first iteratively learns the training tasks which derives the initial parameter $\Theta^0$. A task inference module is established based on $F(\cdot, \Theta^0)$. During the dynamic model adaptation, the task inference module identifies historical data that might contain recurring patterns according to the current input. The forecasting model is then adapted based on the identified and the latest data. Before starting the next model adaptation, the forecasting model learns the latest labeled data using a single gradient step.
	
\begin{figure}
		\centering
		\includegraphics[width=0.45\textwidth]{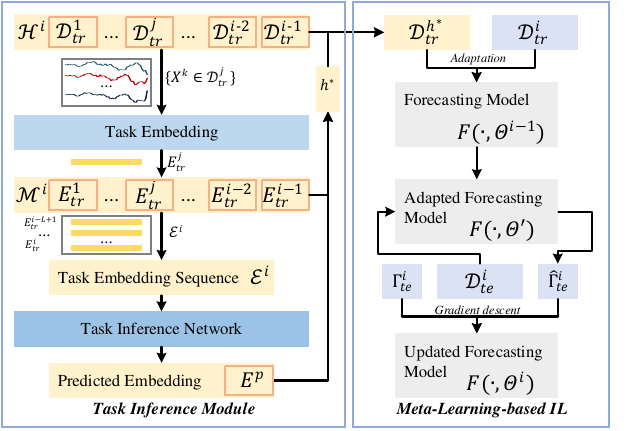}
		\caption{The overview of MetaDA's working pipeline in the $i$ th model adaptation. A set of historical data $\mathcal{D}_{tr}^{h^*}$ is first selected by the task inference module. Model adaptation is then conducted based on $\mathcal{D}_{tr}^{h^*}$ and the latest labeled data $\mathcal{D}_{tr}^{i}$. The forecasting model subsequently predicts stock trend on $\mathcal{D}_{te}^{i}$, and learns the data after the arrival of labels $\mathbf{\Gamma}_{te}^i$.}
		\label{overview}
		\vspace{-0.3em}
	\end{figure}

	Both the training and adapting processes are based on \textit{model adaptation}, which can be viewed as the atomic operation of the above procedure. The working pipeline of a single model adaptation is presented in Figure \ref{overview}. The following subsections will introduce the task inference module (\ref{TaskIM}), meta-learning-based IL (\ref{MetaIL}), and the details of training-adapting procedure (\ref{TrainingAP}).
	
	\subsection{Task Inference Module}
	\label{TaskIM}
	As shown in Figure \ref{overview}, the task inference module can be further divided into two key components: task embedding and task inference network. For convenience, the two major components are introduced in the scenario of \textit{$i$ th model adaptation}. Task inference module aims to select a proper task $D^{h^*}_{tr}$ from historical data $\mathcal{H}^i$, preparing the model $F(\cdot; \Theta)$ for the predictive concept drifts. 
	
	\vspace{3.5pt}
	\noindent\textbf{Task Embedding.} The objective of the task embedding is to project the training sets $\{\mathcal{D}_{tr}^j \}_{j=1}^{i}$ into learnable vectors $\{\mathbf{E}_{tr}^j\}_{j=1}^{i}$, namely the task-level embeddings.
	The corresponding stock features $\{\mathbf{X}|\mathbf{X} \in \mathcal{D}_{tr}^j \}_{j=1}^{i}$ are first encoded into hidden features, which are denoted as sample-level embeddings. The sample-level embeddings are extracted as:
	\begin{equation}
		\label{sampleEncode}
		\mathbf{e}^k = \rm{Encoder}(\mathbf{X}^k),
	\end{equation}
	where $\mathbf{e}^k=\{\mathbf{s}^i \in \mathbb{R}^{1\times q}\}_{i=1}^{n_k}$ denotes the sample-level embeddings corresponding to $\mathbf{X}^k$ in the training sets, and $n_k$ is the number of samples at date $k$. It should be noted that the number of samples (stock data) at each date is variable due to stock suspensions and other unforeseen circumstances, namely $\forall k,l: n_k \not\equiv n_l$. 
	
	After encoding the stock data, the ensuing problem is how to extract effective task-level embedding $\mathbf{E}^j_{tr}$ from sample-level embeddings. As each $\mathbf{e}^k=\{\mathbf{s}^i\}_{i=1}^{n_k}$ contains an uncertain number of sample-level embeddings, the extractor should be independent of sample numbers. In this paper, we introduce an attention-based aggregating layer for task embedding, where sample-level embeddings are aggregated with a weighted average process.
	\begin{equation}
		\mathbf{E}^j_{tr} = \sum_{k=T_{j-1}}^{T_j - 1} \sum_{i=1}^{n_k} \alpha^i (\mathbf{s}^i\mathbf{V}_1 + \boldsymbol{\epsilon}),
	\end{equation}
	where $\mathbf{V}_1 \in \mathbb{R}^{q\times q}$ and $\boldsymbol{\epsilon} \in \mathbb{R}^{1\times q}$ are the parameters of linear projection. The aggregating weights $\alpha$ are learned by the embedding model and processed with softmax to make it invariant to the sample numbers of tasks:
	\begin{equation}
		\label{atta}
		\alpha^i = \frac{\exp \{\tanh(\mathbf{s}^i \mathbf{V}_2)\mathbf{v}_3\}}{\sum_{k=T_{j-1}}^{T_j - 1} \sum_{i=1}^{n_k} \exp \{\tanh(\mathbf{s}^i \mathbf{V}_2)\mathbf{v}_3\}} .
	\end{equation}
	In Eq. \ref{atta}, $\mathbf{V}_2 \in \mathbb{R}^{q\times p}$ and $\mathbf{v}_3 \in \mathbb{R}^{p \times 1}$. The above equation corresponds to a version of attention mechanism \cite{atticlr}, which could discover similarities among the sample-level embeddings. The learnable parameters of task embedding are represented by $\Pi_1=\{\mathbf{V}_1,\mathbf{V}_2,\mathbf{v}_3\}$
	
	
	\vspace{3.5pt}
	\noindent\textbf{Task Inference Network.} The previous study indicates the existence of predictable concept drift in the stock data. We therefore assume that the patterns of tasks are partially predictable: the pattern of upcoming task can be inferred from a sequence of historical tasks. The aforementioned task embeddings are first organized into embedding sequences. In the $i$ th model adaptation, the task embedding sequence is formulated as follows.
	\begin{equation}
		\label{taskembed}
		\mathbf{\mathcal{E}}^i = \{\mathbf{E}_{tr}^{i-L+1},\cdots,\mathbf{E}_{tr}^{i-1},\mathbf{E}_{tr}^{i}\},
	\end{equation}
	where $\mathbf{\mathcal{E}}^i$ is the task embedding sequence for predicting the upcoming task, and $L$ is the size of lookback window. Meanwhile, the embeddings of all the previous tasks are collected into a memory list, which is denoted as follows
	\begin{equation}
		\mathbf{\mathcal{M}}^i = \{\mathbf{E}^1_{tr},\cdots ,\mathbf{E}_{tr}^{i-2},\mathbf{E}_{tr}^{i-1}\},
	\end{equation}
	where $\mathbf{\mathcal{M}}^i$ denotes the memory list of $i$ th model adaptation. 
	
	To predict the embedding of the upcoming task, the embedding sequence is fed into the task inference network, which derives an estimated embedding $\mathbf{E}_p^i$. 
	\begin{equation}
		\mathbf{E}_p^i = F_{tin}(\mathbf{\mathcal{E}}^i, \Pi_2)
	\end{equation}
	where $F_{tin}(\cdot, \Pi_2)$ denotes the task inference network with model parameter $\Pi_2$. The predicted embedding is utilized to select a historical task that contains the task-level pattern most similar to the predicted pattern. The index of selected historical task is denoted as $h^*$ and determined as:
	\begin{equation}
		\min_{h^*} \Vert\mathbf{E}^{h^*}_{tr} - \mathbf{E}_p^i\Vert_F^2, \quad
		s.t.\, \mathbf{E}^{h^*}_{tr} \in  \mathbf{\mathcal{M}}^i.
	\end{equation}
	The reliability of $h^*$ is further estimated and the inference results with low reliability will be ignored. $\mathbf{E}^{h^*}_{tr}$ is included for model adaptation only if the similarity between $\mathbf{E}^{i}_{p}$ and $\mathbf{E}^{i}_{tr}$ is in $\kappa$ th percentile.
	
	\subsection{Meta-Learning-based IL} 
	\label{MetaIL}
	In this paper, the incremental learning of stock trends is organized in the paradigm of meta-learning \cite{Dadpt}. Meta-Learning-based IL contains a forecasting model $F(\cdot; \Theta)$ which incorporates a data adapter $f_{da}(\cdot; \theta_{da})$, a label adapter $f_{la}(\cdot; \theta_{la})$ and a forecaster $f(\cdot; \theta_{f})$. $\theta_{da}$ and $\theta_{la}$ are the parameters of the data adapter and label adapter correspondingly, while $\theta_{f}$ is the parameter of the forecaster. $\Theta$ is the collection of above parameters with $\Theta = \{\theta_{da}, \theta_{la}, \theta_{f}\}$.
	
	In the $i$ th model adaptation, the forecaster is first updated with $\mathcal{D}^{i}_{ada}$. If $\mathcal{D}^{h^*}_{tr}$ is included for adaptation, $\mathcal{D}^{i}_{ada}=\{(\mathbf{X}, \mathbf{\Gamma}) \in \mathcal{D}^{i}_{tr}\cup \mathcal{D}^{h^*}_{tr} \}$. Otherwise, $\mathcal{D}^{i}_{ada}=\mathcal{D}^{i}_{tr}$. The initial parameters of forecasting model in the $i$ th model adaptation are updated parameters of $i-1$ th model adaptation, denoted as $\Theta^{i-1}=\{\theta_{da}^{i-1}, \theta_{la}^{i-1}, \theta_{f}^{i-1}\}$. The stock data are first transformed with the data adapter and label adapter for the alleviation of concept drift. Suppose there are data $\mathbf{X} \in \mathbb{R}^{n \times d}$ and corresponding labels $\mathbf{\Gamma} \in \mathbb{R}^{n \times 1}$,
	\begin{equation}
		\begin{aligned}
			&\tilde{\mathbf{X}}_{(\theta_{da})} = f_{da}(\mathbf{X}; \theta_{da}) := \mathbf{X} + \sum^{N}_{i} \boldsymbol{\beta}_i (\mathbf{X}\mathbf{W}_i + \mathbf{b}_i), \\
			&\tilde{\mathbf{\Gamma}}_{(\theta_{la})} = f_{la}(\mathbf{\Gamma}; \theta_{la}) := \sum^{N}_{i} \boldsymbol{\beta}_i (\mathbf{\Gamma}\cdot h_i + z_i),
		\end{aligned}
	\end{equation}
	where $\mathbf{W}_i \in \mathbb{R}^{d \times d}$ and $\mathbf{b}_i$ are the parameters of $i$ th linear projection for data adaptation. $h_i \in \mathbb{R}$ and $z_i$ are the parameters of $i$ th linear projection for label adapter, while $N$ is the number of projection. $\boldsymbol{\beta}_i \in \mathbb{R}^{n \times n}$ is the weight matrix of $i$ th projection.
	\begin{equation}
		\boldsymbol{\beta}_i = \text{diag}(\beta_{(i,1)}, \beta_{(i,2)}, \cdots, \beta_{(i,n)}),
	\end{equation} 
	where $\beta_{(i,j)}$ represents the weight of $i$ th projection with respect to the $j$ th sample:
	\begin{equation}
		\beta_{(i,j)} = \frac{\exp(\cos<\mathbf{x}^j, \mathbf{p}^i>/\omega^2)}{\sum_{i=1}^N \exp(\cos<\mathbf{x}^j, \mathbf{p}^i>/\omega^2)},
	\end{equation}
	where $\cos<\cdot, \cdot>$ denotes the cosine similarity, and $\mathbf{p}^i$ is a learnable parameter corresponding to $i$ th projection. Meanwhile, $\mathbf{x}^j$ denotes the $j$ th sample within $\mathbf{X}$.  $\omega$ is the softmax temperature. In general, we have $\theta_{da}=\{\mathbf{\Gamma}_i, \mathbf{b}_i, \mathbf{p}^i\}_{i=1}^N$ and $\theta_{la}=\{h_j, z_j, \mathbf{p}^j\}_{j=1}^{N}$.
	
	The parameter of forecaster $\theta_{f}^{i-1}$ is then updated with a single gradient step using $\tilde{\mathcal{D}}^{i}_{ada}=\{(\tilde{\mathbf{X}}_{(\theta_{da}^{i-1})}, \tilde{\mathbf{\Gamma}}_{(\theta_{la}^{i-1})})|(\mathbf{X}, \mathbf{\Gamma}) \in \mathcal{D}^{i}_{ada} \}$. In this paper, the meta-learning process has optimized $\Theta$ for quick adaptation to the upcoming tasks. A single gradient step is therefore enough to boost the performance. 
	\begin{equation}
		\label{oadapt}
		\theta_{f}' \gets \theta_{f}^{i-1} - \eta \nabla_{\theta_{f}} \textbf{L}_{ada}({\tilde{\mathcal{D}}}^{i}_{ada};\theta_f^{i-1}),
	\end{equation}
	where $\eta$ is the learning rate of forecaster, and $\textbf{L}_{ada}$ is the loss function for model adaptation. The parameters of adapted forecasting model are denoted as $\Theta'=\{\theta_{da}^{i-1},\theta_{la}^{i-1},\theta_{f}'\}$, which are utilized for subsequent prediction on $\mathcal{D}^{i}_{te}$.  Given $\mathbf{X} \in \mathcal{D}^{i}_{te}$, the stock trend is predicted as following:
	\begin{equation}
		\hat{\mathbf{\Gamma}} = F(\mathbf{X}; \Theta')  = f^{-1}_{la}(f(\tilde{\mathbf{X}}_{(\theta_{da}^{i-1})};\theta_{f}'); \theta_{la}^{i-1}), 
	\end{equation}
	where $f^{-1}_{la}= \sum^{N}_{i} \boldsymbol{\beta}_i \mathbf{B}_i (\mathbf{\Gamma} - z_i)/h_i$ is the inverse of label adapter that projects the results back into the original space.
	
	After the arrival of true labels, $\mathcal{D}^{i}_{te}$ and its corresponding labels form $\mathcal{D}^{i+1}_{tr}$. An online gradient step is carried out to update the forecasting model, deriving the initial parameters of the $i+1$ th model adaptation (namely $\Theta^{i}$). 
	\begin{equation}
		\label{ogd}
		\begin{aligned}
			&\theta_{f}^i \gets \theta_{f}^{i-1} - \eta \nabla_{\theta_{f}} \textbf{L}_{ogd}(\mathcal{D}^{i+1}_{tr};\Theta'), \\
			&\theta_{da}^i \gets \theta_{da}^{i-1} - \eta_a \nabla_{\theta_{da}} \textbf{L}_{ogd}(\mathcal{D}^{i+1}_{tr};\Theta'), \\
			& \theta_{la}^i \gets \theta_{la}^{i-1} - \eta_a \nabla_{\theta_{la}} \textbf{L}_{ogd}(\mathcal{D}^{i+1}_{tr};\Theta'),
		\end{aligned}
	\end{equation}
	where $\textbf{L}_{ogd}$ is the loss function for online gradient descent, and $\eta_a$ is the learning rate of adapters.
	
	\subsection{Training-Adapting Procedure}
	\label{TrainingAP}
	In this subsection, the detailed training-adapting procedure of MetaDA is presented, which could be divided into three stages: training of forecasting model, training of task inference module, and adaptation.
	
	\vspace{3.5pt}
	\noindent \textbf{Training of Forecasting Model.} In this stage, we aim to obtain the initial parameters $\Theta^0$ that can efficiently adapt to the upcoming tasks. The training set and validation set are split into tasks according to the adaptation interval $T_{ada}$, deriving a training sequence $\mathcal{T}_{tr}$ and a validation sequence $\mathcal{T}_{val}$. Each task $\mathcal{T}_{j}$ within $\mathcal{T}_{tr}$ and $\mathcal{T}_{val}$ satisfies $\mathcal{T}_{j} =\{\mathcal{D}^{j}_{tr}, \mathcal{D}^{j}_{te}, \mathcal{H}_{j}\}$, same as the above sections. Within a training epoch, the forecasting model performs the model adaptation process on each task $\mathcal{T}_j \in \mathcal{T}_{tr}$. It should be noted that only the latest data are used for model adaptation in this stage, namely $\mathcal{D}^{i}_{ada}=\mathcal{D}^{i}_{tr}$. For the training of forecasting model, the loss function for model adaptation and online learning is defined as:
	\begin{equation}
		\begin{aligned}
			&\mathbf{L}_{ada}(\tilde{\mathcal{D}}; \theta_f) = \mathbf{L}_{mse}(f(\mathbf{X};\theta_f),\mathbf{\Gamma} |\tilde{\mathcal{D}}) \\
			&\mathbf{L}_{ogd}(\mathcal{D}; \Theta) =  \mathbf{L}_{mse}(F(\mathbf{X};\Theta),\mathbf{\Gamma} |\mathcal{D}) + R_{la},
		\end{aligned}	
	\end{equation}
	where $\mathbf{L}_{mse}(F(\mathbf{X};\Theta),\mathbf{\Gamma} |\mathcal{D})$ denotes the mean square error loss $\sum_{(\mathbf{X}, \mathbf{\Gamma})\in\mathcal{D}} \Vert F(\mathbf{X};\Theta) - \mathbf{\Gamma} \Vert_F^2 /N_{\mathcal{D}}$, and $N_{\mathcal{D}}$ is the number of samples within $\mathcal{D}$. 
	$R_{la} = \frac{1}{2}\sum_{ \mathbf{\Gamma}\in\mathcal{D}} \Vert \mathbf{\Gamma} - \tilde{\mathbf{\Gamma}} \Vert_F^2 / N_{\mathcal{D}}$ is a regularizer of label adaptation.
	In a single epoch, $F(\cdot;\Theta)$ learns all the tasks within $\mathcal{T}_{tr}$ and the corresponding parameter is saved as $\Theta_{tmp}$. The forecasting model is then evaluated on $\mathcal{T}_{val}$ following the same model adaptation process. Finally, the saved parameter $\Theta_{tmp}$ is restored, and a new learning epoch on the $\mathcal{T}_{tr}$ is started. The training stops when the performance of $F(\cdot;\Theta)$ decreases on $\mathcal{T}_{val}$ for several continuous epochs, deriving the initial parameters $\Theta^0$.
	
	\vspace{3.5pt}
	\noindent \textbf{Training of Task Inference Module.} The task inference module is subsequently trained based on $\Theta^0$. We exploit the encoding part of $f(\cdot;\theta^0_f)$, which denoted as $f_e(\cdot;\theta^0_f)$, for the sample-level encoding. 
	\begin{equation}
		\mathbf{e}^k = f_{e}(\tilde{\mathbf{X}}^k_{(\theta_{da}^0)}; \theta^0_f).
	\end{equation}
	The parameters of the sample-level encoder remain unchanged. The task embeddings are organized into aforementioned task embedding sequences defined in Eq. \ref{taskembed}. The learnable parameters of the task inference module (namely $\Pi_1$ and $\Pi_2$) are optimized collaboratively. Given a task embedding sequence $\mathcal{E}^i$ of $\mathcal{T}_i$, the task inference network $F_{tin}(\cdot;\Pi_2)$ will output a predicted embedding $\mathbf{E}_p^i$. We hope the predicted embedding $\mathbf{E}_p^i$ has a significant similarity to the embedding of upcoming data $\mathcal{D}^i_{te}$ and has less similarities to the other task embeddings. A triplet loss function is therefore introduced, which is defined as follows.
	\begin{equation}
		\begin{aligned}
			& \mathbf{L}_{tri}(\mathbf{E}_p, \mathbf{E}_t, \mathbf{E}_n) = \\ & \max(\Vert \mathbf{E}_p -\mathbf{E}_t \Vert_2 - \Vert \mathbf{E}_p -\mathbf{E}_n \Vert_2 + \gamma, 0),
		\end{aligned}	
	\end{equation}
	where $\mathbf{E}_n$ denotes the negative embedding and $\mathbf{E}_t$ is the target embedding, namely the embedding of $\mathcal{D}^i_{te}$. $\gamma$ is the margin of triplet loss. The task embeddings of $\mathcal{T}_{tr}$ are correspondingly organized into triplets, from which we derive the training triplets $\mathbf{Trip}_{tr}$.
	\begin{equation}
		\begin{aligned}
			&\mathbf{Trip}_{tr} = \{\mathbf{trip}^i \}_{i=L}^{N_{tr}}, \\
			&\mathbf{trip}^i =  \{(\mathbf{E}_p^i, \mathbf{E}_{tr}^{i+1}, \mathbf{E}_{tr}^{k})| \mathbf{E}_{tr}^{k} \in \mathcal{E}^i_{neg} \}, \\
			& \mathcal{E}^i_{neg} = \{\mathbf{E}_{tr}^{k}|k\neq i, k\neq i+1\}_{k=1}^{N_{tr}},
		\end{aligned}
	\end{equation}
	where $N_{tr}$ is the number of tasks in the $\mathcal{T}_{tr}$, and $\mathbf{trip}^i$ denotes the triplets corresponding to the $i$ th task. $\mathbf{E}_{tr}^{i+1}$ is equal to $\mathbf{E}_{te}^{i}$ and therefore used as the target embedding of $\mathbf{E}_p^{i}$, while the rest embeddings except $\mathbf{E}_{tr}^{i}$ serve as negative embeddings.
	
	Within a single epoch, triplets of each task are learned with a learning rate of $\eta_{t}$ and the learned parameters are saved as $\{\Pi_1^{tmp},\Pi_2^{tmp}\}$. The task inference model is then evaluated on the validation set based on $\{\Pi_1^{tmp},\Pi_2^{tmp}\}$ and $\Theta^0$. Different from the training of forecasting model, $\mathcal{D}_{tr}^{h^*}$ is introduced for model adaptation. When the evaluation performance on $\mathcal{T}_{val}$ decreases for several continuous epochs, the training process of task inference module stops.
	
	\vspace{3.5pt}
	\noindent\textbf{Adapting.} After the training stage, the forecasting model first performs the model adaptation process on all the tasks within $\mathcal{T}_{val}$ in sequence. When forecasting the trends of streaming stock data, MetaDA conducts model adaptation periodically following the process described in section \ref{TaskIM} and \ref{MetaIL}.
	
	\section{Experiments}
	In this section, the proposed MetaDA is evaluated with several experiments, aiming to answer the following research questions. The implementation is publicly available\footnote{provided in \textit{Supplementary Material} for anonymity}. 
	
	\noindent\hangindent=2em\textbf{Q1.} How does MetaDA perform compared with state-of-the-art methods?
	
	\noindent\textbf{Q2.} Are the proposed components in MetaDA effective?
	
	\noindent\textbf{Q3.} Can MetaDA learn the incremental data efficiently?
	
	\begin{table*}[!ht]
		\centering
		\footnotesize
		\renewcommand{\arraystretch}{1.05}
		\begin{tabular}{p{1.5cm}||p{1.4cm}||p{0.72cm}p{0.72cm}p{0.72cm}p{0.72cm}p{0.72cm}p{0.75cm}|p{0.72cm}p{0.72cm}p{0.72cm}p{0.72cm}p{0.72cm}p{0.72cm}p{0.72cm}}
			\toprule
			\multicolumn{1}{c||}{\multirow{2}{*}{Forecaster}}  & \multicolumn{1}{c||}{\multirow{2}{*}{Method}} & \multicolumn{6}{c|}{CSI-300}                                                                               & \multicolumn{6}{c}{CSI-500}                                                                               \\
			& \multicolumn{1}{c||}{}                        & IC              & ICIR            & RIC             & RICIR           & eAR             & eARIR           & IC              & ICIR            & RIC             & RICIR           & eAR             & eARIR           \\ \midrule
			\multicolumn{1}{c||}{\multirow{6}{*}{LSTM}}        & DDG-DA                                      & 0.0452          & 0.3610          & 0.0546          & 0.4460          & 0.1193          & 1.6619          & 0.0530          & \ul{0.4832}    & 0.0670          & 0.6610          & 0.1204          & 2.5623          \\
			& IL                                          & 0.0484          & 0.3476          & 0.0633          & 0.4663          & 0.1563          & 2.2458          & 0.0529          & 0.4518          & 0.0719          & 0.6790          & 0.1380          & 2.8764          \\
			& C-MAML                                      & 0.0473          & 0.3641          & 0.0587          & 0.4577          & 0.1449          & 2.0385          & 0.0473          & 0.4019          & 0.0647          & 0.6120          & 0.1021          & 2.1079          \\
			& MetaCoG                                     & 0.0348          & 0.2417          & 0.0497          & 0.3532          & 0.0891          & 0.0891          & 0.0523          & 0.4658          & 0.0662          & 0.6590          & 0.1192          & 2.6571          \\
			& DoubleAda                                   & {\ul{0.0520}}    & {\ul{0.3891}}    & {\ul{0.0651}}    & {\ul{0.4969}}&\ul{0.1577}    & {\ul{0.1577}}    & {\ul{0.0554}}    & 0.4721          & {\ul{0.0737}}    & {\ul{0.7035}}    & {\ul{0.1422}}    & {\ul{2.9799}}    \\
			& \textbf{MetaDA}                             & \textbf{0.0525} & \textbf{0.4010} & \textbf{0.0657} & \textbf{0.5179} & \textbf{0.1661} & \textbf{2.4854} & \textbf{0.0577} & \textbf{0.5165} & \textbf{0.0745} & \textbf{0.7418} & \textbf{0.1474} & \textbf{3.1736} \\ \midrule
			\multicolumn{1}{c||}{\multirow{6}{*}{GRU}}         & DDG-DA                                      & 0.0532          & 0.4027          & 0.0604          & 0.4722          & 0.0961          & 1.3743          & 0.0508          & 0.4668          & 0.0640          & 0.6266          & 0.1078          & 2.4081          \\
			& IL                                          & 0.0536          & 0.3815          & 0.0664          & 0.4968          & 0.1641          & 2.3646          & 0.0554          & 0.4716          & 0.0730          & 0.6891          & 0.1372          & 2.8210          \\
			& C-MAML                                      & 0.0547          & 0.4135          & 0.0643          & 0.4940          & 0.1670          & 2.4145          & 0.0532          & 0.4497          & 0.0710          & 0.6725          & 0.1214          & 2.5133          \\
			& MetaCoG                                     & 0.0461          & 0.3365          & 0.0583          & 0.4430          & 0.1360          & 1.9555          & 0.0530          & 0.4883          & 0.0669          & 0.6777          & 0.1153          & 2.6131          \\
			& DoubleAda                                   & {\ul{0.0570}}    & {\ul{0.4186}}    & \textbf{0.0686} & {\ul{0.5198}}    & {\ul{0.1733}} & \ul{2.4994}    & {\ul{0.0584}}    & {\ul{0.5070}}    & {\ul{0.0755}}    & {\ul{0.7331}}    & {\ul{0.1470}}    & {\ul{3.0941}}    \\
			& \textbf{MetaDA}                             & \textbf{0.0584} & \textbf{0.4400} & {\ul{0.0683}}    & \textbf{0.5372} & \textbf{0.1792} & \textbf{2.7146} & \textbf{0.0606} & \textbf{0.5361} & \textbf{0.0757} & \textbf{0.7412} & \textbf{0.1506} & \textbf{3.1794} \\ \midrule
			\multirow{6}{*}{Transformer} & DDG-DA                                      & 0.0161          & 0.1043          & 0.0353          & 0.2516          & 0.0301          & 0.4360          & 0.0119          & 0.0851          & 0.0372          & 0.3025          & 0.0244          & 0.4472          \\
			& IL                                          & 0.0333          & 0.2269          & 0.0489          & 0.3493          & 0.0970          & 1.3328          & 0.0395          & 0.3155          & 0.0614          & 0.5401          & {\ul{0.1000}}    & {\ul{1.8332}}    \\
			& C-MAML                                      & 0.0346          & 0.2422          & 0.0495          & 0.3637          & 0.0992          & 1.3547          & 0.0415          & 0.3217          & 0.0603          & 0.5279          & 0.0806          & 1.4656          \\
			& MetaCoG                                     & 0.0163          & 0.1015          & 0.0229          & 0.1390          & 0.0228          & 0.2473          & 0.0110          & 0.0888          & 0.0230          & 0.1909          & 0.0275          & 0.5949          \\
			& DoubleAda                                   & {\ul{0.0359}}    & {\ul{0.2437}}    & {\ul{0.0510}}    & {\ul{0.3660}}    & {\ul{0.1146}}  &\ul{1.4721}   & {\ul{0.0490}}    & {\ul{0.3853}}    & \textbf{0.0659} & {\ul{0.5664}}    & 0.0866          & 1.5697          \\
			& \textbf{MetaDA}                             & \textbf{0.0430} & \textbf{0.3061} & \textbf{0.0555} & \textbf{0.4216} & \textbf{0.1270} & \textbf{1.9027} & \textbf{0.0552} & \textbf{0.4404} & {\ul{0.0652}}    & \textbf{0.5732} & \textbf{0.1031} & \textbf{1.8928} \\ \bottomrule
		\end{tabular}
		\caption{The overall performance comparison on the CSI-300 and CSI-500 datasets, where IC, ICIR, RIC, RICIR, eAR, and eARIR are utilized as metrics. The best results are highlighted in bold, and the second-best results are underlined.}
		\label{compart}
		\vspace{-0.3em}
	\end{table*}

	\subsection{Experimental Settings}
	\textbf{Datasets.} Two widely-used stock datasets are introduced for evaluation, namely CSI-300 \cite{csi300} and CSI-500 \cite{REST}. Both datasets are collected in the Chinese A-share market and contain the stock data from 01/01/2008 to 31/07/2020. CSI-300 consists of the 300 largest stocks, while CSI-500 consists of the 500 largest remaining stocks. The datasets are split into training sets (01/01/2008 to 31/12/2014), validation sets (01/01/2015 to 31/12/2016), and testing sets (01/01/2017 to 31/07/2020).
	
	The stock feature of Alpha360 in Qlib \cite{qlib} is utilized for stock forecasting. On each trading day, Alpha360 records 6 indicators in the past 60 days, including opening price, closing price, highest price, lowest price, trading volume, and volume-weighted average price.
	
	\vspace{3.5pt}
	\noindent\textbf{Metrics.} Four widely used metrics are introduced for the evaluation of predicted stock trends, namely Information Coefficient (IC), IC information ratio (ICIR), rank IC, and rank ICIR. IC at date $t$ is defined as the Pearson correlation between the estimated labels and the raw labels. The average IC across all the testing dates is reported in this paper. ICIR is defined as $\mathbb{E}(IC)/\sigma(IC)$, where $\sigma(\cdot)$ denotes standard deviation. Rank IC (RIC) and rank ICIR (RICIR) are calculated in a similar way, where the ranks of labels are used.
	Portfolio metrics are also included for evaluation, including the excess annualized return (eAR) and the excess AR’s information ratio (eARIR), where TopkDropout strategy \cite{qlib} is utilized. Friedman test \cite{freidman} has been introduced for quantitative analysis across the above datasets and metrics. For all the aforementioned metrics, higher values represent better performance.

	\vspace{3.5pt}
	\noindent\textbf{Baselines.} Several state-of-the-art model-agnostic retraining methods are introduced for comparison:
	
	\begin{itemize}[leftmargin=8pt, labelsep = 3pt, topsep=0pt, parsep=0pt]
		\item\textbf{DDG-DA} \cite{DDGDA}: This rolling retraining method re-weights the historical data according to the predicted data distribution, and periodically retrains the forecasting model with the re-weighted data.
		
		\item\textbf{IL} \cite{Dadpt}: IL, short for incremental learning, periodically updates the model with the latest data using online gradient descent. 
		
		\item\textbf{C-MAML} \cite{CMAML}: Based on model-agnostic meta-learning (MAML), this method first trains a general model across various training tasks, and performs model adaptation with the latest data periodically.
		
		\item\textbf{MetaCoG} \cite{MetaCoG}: Introducing a continuously updated per-parameter mask, this MAML-based method selects task-specific parameters according to the context.
		
		\item\textbf{DoubleAda} \cite{Dadpt}: This MAML-based method performs additional data and label adaptations to alleviate the impact of concept drifts, which uses the latest data exclusively for model adaptation.
		
		\item\textbf{MetaDA}: Following the paradigm of MAML, the proposed method performs model adaptation with both the latest data and the selected historical data.
	\end{itemize}
	
	\noindent\textbf{Implementation.} The time interval $T_{ada}$ between two model adaptation is 15 trading days. Adam optimizer is utilized for training, where $\eta=\eta_{t}=0.001$ and $\eta_a=0.01$. During all the following experiments, a single-layer GRU \cite{GRU} is used as the task inference network with $\kappa$ fixed at 80, $\gamma$ fixed at 1 and $N$ fixed at 8. In the following parts, the average results of 5 repeated experiments are reported.
	
	
	
	\begin{table*}[]
		\centering
		\footnotesize
		\renewcommand{\arraystretch}{1.05}
		\begin{tabular}{l||p{0.75cm}p{0.75cm}p{0.75cm}p{0.75cm}|p{0.75cm}p{0.75cm}p{0.75cm}p{0.75cm}|p{0.75cm}p{0.75cm}p{0.75cm}p{0.75cm}}
			\toprule
			\multicolumn{1}{c||}{\multirow{2}{*}{}} & \multicolumn{4}{c|}{LSTM}                                              & \multicolumn{4}{c|}{GRU}                                               & \multicolumn{4}{c}{Transformer}                                       \\
			\multicolumn{1}{c||}{}                  & IC              & ICIR            & RIC         & RICIR       & IC              & ICIR            & RIC         & RICIR       & IC              & ICIR            & RIC         & RICIR       \\ \midrule
			IL/CSI-300                            & 0.0484          & 0.3476          & 0.0633          & 0.4663          & 0.0536          & 0.3815          & 0.0664          & 0.4968          & 0.0333          & 0.2269          & 0.0489          & 0.3493          \\
			MetaIL/CSI-300                        & 0.0517          & 0.3927          & 0.0644          & 0.5019          & 0.0573          & 0.4306          & 0.0678          & 0.5304          & 0.0406          & 0.2863          & 0.0526          & 0.3902          \\
			MetaDA/CSI-300                        & \textbf{0.0525} & \textbf{0.4010} & \textbf{0.0657} & \textbf{0.5179} & \textbf{0.0584} & \textbf{0.4400} & \textbf{0.0683} & \textbf{0.5372} & \textbf{0.0430} & \textbf{0.3061} & \textbf{0.0555} & \textbf{0.4216} \\ \midrule
			IL/CSI-500                            & 0.0529          & 0.4518          & 0.0719          & 0.6790          & 0.0554          & 0.4716          & 0.0730          & 0.6891          & 0.0395          & 0.3155          & 0.0614          & 0.5401          \\
			MetaIL/CSI-500                        & 0.0576          & 0.5081          & 0.0740          & 0.7230          & 0.0597          & 0.5207          & \textbf{0.0759} & 0.7307          & 0.0524          & 0.4265          & \textbf{0.0653} & \textbf{0.5817} \\
			MetaDA/CSI-500                        & \textbf{0.0577} & \textbf{0.5165} & \textbf{0.0745} & \textbf{0.7418} & \textbf{0.0606} & \textbf{0.5361} & 0.0757          & \textbf{0.7412} & \textbf{0.0552} & \textbf{0.4404} & 0.0652          & 0.5732   \\ \bottomrule      
		\end{tabular}
		\caption{The results of ablation experiments on CSI-300 and CSI-500 datasets, where MetaIL is short for Meta-Learning-based IL.}
		\label{abla}
		\vspace{-0.3em}
	\end{table*}
	
	\subsection{Performance Comparison (Q1)}
	To explore the performance of MetaDA across different forecasters, the above methods are evaluated with different forecasters ($f(\cdot, \theta_f)$ for MetaDA). Three widely used deep neural networks with Qlib’s default structure are used, including LSTM \cite{lstm}, GRU \cite{GRU}, and Transformer \cite{Transformer}.
	
\begin{figure}
		\centering
		\includegraphics[width=0.45\textwidth]{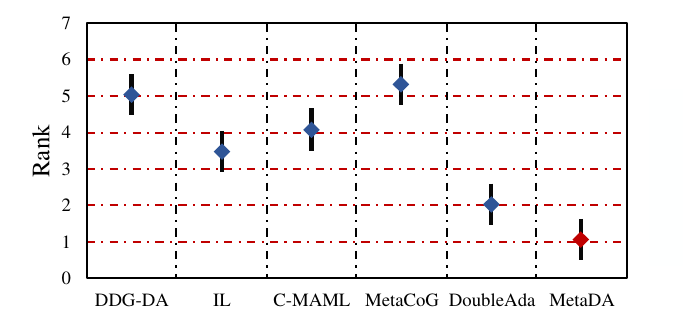}
		\caption{The results of Friedman test at 90\% significance level, which is a general analysis across different datasets, forecasters, and metrics. If the bars of two methods do not overlap, it indicates that they are significantly different.}
		\label{FR1}
		\vspace{-0.3em}
	\end{figure}

	We first carry out comparison experiments where IC, ICIR, RIC, RICIR, eAR, and eARIR are utilized for evaluation. The experimental results are presented in Table \ref{compart}. The proposed MetaDA achieves the best results in most scenarios, indicating our method can enable more accurate stock trend prediction across various datasets and models. Since multiple metrics, forecasters, and datasets have been included, Friedman test is further introduced to evaluate the overall performance across different scenarios. The results of Friedman test at 90\% significance are shown in Figure \ref{FR1}. MetaDA achieves the best average rank among the tested methods, and has significantly outperformed several comparison methods like DDG-DA and C-MAML. When it comes to the latest IL methods, the proposed method still achieves a better average rank with a considerable margin. Compared to DoubleAda, MetaDA has further improved the prediction of stock trends by revising the selected historical data. In this way, MetaDA can effectively update the forecaster with the latest data while alleviating the impacts of catastrophic forgetting. Moreover, the performance of MetaDA can be further boosted by introducing a better forecaster or a better task inference module.
	
	\begin{figure}
		\centering
		\includegraphics[width=0.47\textwidth]{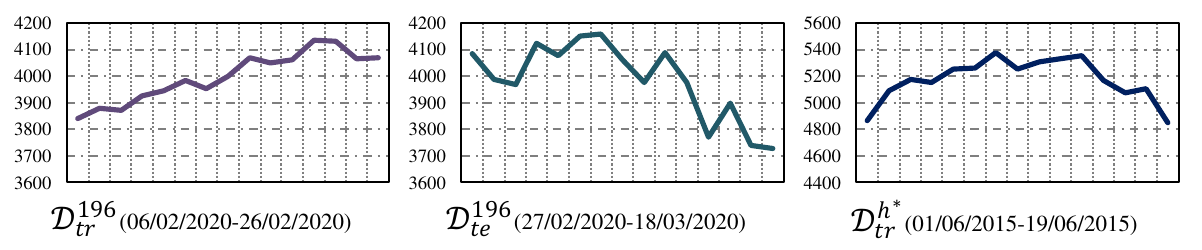}
		\caption{The CSI-300 index (SH000300) of $\mathcal{D}_{tr}^{196}$, $\mathcal{D}_{te}^{196}$, and corresponding $\mathcal{D}_{tr}^{h^*}$, which has a stock price trend similar to that of $\mathcal{D}_{te}^{196}$. With the introduction of $\mathcal{D}_{tr}^{h^*}$, IC has increased by 50 \% on this task.}
		\label{FR2}
		\vspace{-0.3em}
	\end{figure}

	\subsection{Ablation Study (Q2)}
	The results of ablation study on CSI-300 and CSI-500 datasets have been presented in Table \ref{abla}. Three methods are introduced for ablation study, namely IL baseline (IL), MetaIL, and MetaDA. MetaIL denotes meta-learning-based IL and represents the MetaDA without task inference module, which relies exclusively on the latest data for model adaptation. Both MetaDA and MetaIL have achieved performance better than IL, which indicates the effectiveness of meta-learning-based IL framework. Compared with MetaIL, the introduction of dynamic adaptation has boosted the performance of MetaDA in most scenarios. MetaIL achieves better RIC on the CSI-500 datasets with certain forecasters. The methods focusing on recurring patterns (DDG-DA and MetaCoG) have suffered significant performance degradation on the CSI-500 datasets, which indicates that the recurring patterns become less effective. Such performance degradation is occasional, and the introduction of selected historical data has improved performance in most scenarios. A typical example is indicated in Figure \ref{FR2}, where the introduction of historical data has significantly improved the prediction. With $\mathcal{D}_{tr}^{h^*}$, the IC on task $\mathcal{T}_{196}$ has increased by 50\% (from 0.0427 to 0.0660, GRU). In general, the ablation study has proven the effectiveness of the proposed components in MetaDA.
	
	\begin{table}[]
		\centering
		\footnotesize
		\renewcommand{\arraystretch}{1.05}
		\begin{tabular}{lp{1.5cm}p{1.5cm}p{1.5cm}}
			\toprule
			& LSTM   & GRU    & Transformer \\ \midrule
			DDG-DA    & 4974.4 & 4983.2 & 3732.2      \\
			IL        & 1.9    & 1.9    & 4.0         \\
			C-MAML   & 5.4   & 5.7   & 13.4        \\
			MetaCoG   & 3.9   & 4.0   & 8.9        \\
			DoubleAda & 2.9    & 3.5    & 6.1         \\
			\textbf{MetaDA}    & 3.9    & 3.9    & 8.2   \\ \bottomrule     
		\end{tabular}
		\caption{The total time cost (in seconds) of model adaptation/retraining on the CSI-300 dataset.}
		\label{adatime}
		\vspace{-0.3em}
	\end{table}
	\subsection{Model Adaptation Efficiency (Q3)}
	Higher adaptation efficiency means that the models can be frequently adjusted on a larger scale, which could potentially improve stock trend forecasting. The efficiency of model adaptation is therefore evaluated, with the empirical time costs presented in Table \ref{adatime}. The efficiency of MetaDA is comparable with the state-of-the-art methods. As only the selected historical data is added to the model adaptation, the efficiency degradation of MetaDA is limited compared to the baselines of incremental learning. Meanwhile, the proposed method is significantly faster than the rolling retraining method. The above results have indicated that MetaDA has achieved considerable adaptation efficiency.
	
	\section{Conclusion}
	In this paper, meta-learning with dynamic adaptation (MetaDA) is proposed for the incremental learning of stock trends. MetaDA is the first incremental learning method that considers both the predictable and unpredictable concept drifts. This work has provided two insights for alleviating the impact of stock concept drifts. Firstly, dynamic adaptation is incorporated into the framework of meta-learning-based IL, which utilizes the latest and the historical data. Secondly, we propose a task inference module to identify the historical data that might contain recurring patterns. On the real-world datasets, MetaDA has achieved state-of-the-art performance with considerable efficiency. In the future, we are going to incorporate side information such as news into the incremental learning framework for stock trend forecasting.
	
	\appendix
	
	\section*{Ethical Statement}
	
	There are no ethical issues.
	
	\section*{Acknowledgments}

	\bibliographystyle{named}
	\bibliography{ijcai24}
	
\end{document}